\PassOptionsToPackage{dvipsnames}{xcolor}
\documentclass[sigconf]{acmart}

\usepackage{multirow}
\usepackage{balance}
\usepackage{tabularx}
\usepackage{graphicx}
\usepackage{hhline}
\usepackage{pifont}
\usepackage{libertine}
\usepackage[dvipsnames]{xcolor}
\usepackage{appendix}
\usepackage{float}

\usepackage[symbol]{footmisc}
\newcommand{\cmark}{\text{\ding{51}}}
\newcommand{\xmark}{\text{\ding{55}}}

\AtBeginDocument{%
  \providecommand\BibTeX{{%
    \normalfont B\kern-0.5em{\scshape i\kern-0.25em b}\kern-0.8em\TeX}}}

\copyrightyear{2023}
\acmYear{2023}
\setcopyright{acmlicensed}\acmConference[ICTIR '23]{Proceedings of the 2023 ACM SIGIR International Conference on the Theory of Information Retrieval}{July 23, 2023}{Taipei, Taiwan}
\acmBooktitle{Proceedings of the 2023 ACM SIGIR International Conference on the Theory of Information Retrieval (ICTIR '23), July 23, 2023, Taipei, Taiwan}
\acmPrice{15.00}
\acmDOI{10.1145/3578337.3605143}
\acmISBN{979-8-4007-0073-6/23/07}

\begin{document}

\title{Zero-shot Query Reformulation for Conversational Search}


\author{Dayu Yang}
\affiliation{%
  \institution{Institute for Financial Services Analytics \\ University of Delaware}
    \city{Newark}
    \state{DE}
    \country{USA}}
\email{dayu@udel.edu}

\author{Yue Zhang}

\affiliation{%
  \institution{Department of Electrical \& Computer Engineering \\ University of Delaware}
  \city{Newark}
  \state{DE}
  \country{USA}}
\email{zhangyue@udel.edu}

\author{Hui Fang}

\affiliation{%
  \institution{Department of Electrical \& Computer Engineering \\ University of Delaware}
  \city{Newark}
  \state{DE}
  \country{USA}}
\email{hfang@udel.edu}

\renewcommand{\shortauthors}{Dayu Yang, Yue Zhang, \& Hui Fang}

\begin{abstract}


As the popularity of voice assistants continues to surge, conversational search has gained increased attention in Information Retrieval. However, data sparsity issues in conversational search significantly hinder the progress of supervised conversational search methods. Consequently, researchers are focusing more on zero-shot conversational search approaches. Nevertheless, existing zero-shot methods face three primary limitations: they are not universally applicable to all retrievers, their effectiveness lacks sufficient explainability, and they struggle to resolve common conversational ambiguities caused by omission.
To address these limitations, we introduce a novel Zero-shot Query Reformulation (or Query Rewriting) (ZeQR) framework that reformulates queries based on previous dialogue contexts without requiring supervision from conversational search data. Specifically, our framework utilizes language models designed for machine reading comprehension tasks to explicitly resolve two common ambiguities: coreference and omission, in raw queries. In comparison to existing zero-shot methods, our approach is universally applicable to any retriever without additional adaptation or indexing. It also provides greater explainability and effectively enhances query intent understanding because ambiguities are explicitly and proactively resolved.
Through extensive experiments on four TREC conversational datasets, we demonstrate the effectiveness of our method, which consistently outperforms state-of-the-art baselines.


\end{abstract}





\maketitle

\section{Introduction}


Conversational search has gained significant attention in recent years due to the growing use of conversational agents, such as chatbots and voice assistants. These agents have transformed the way users engage with search engines by shifting from keyword-based search to conversational modes. One of the main challenges in developing effective conversational search systems is data sparsity. For  supervised conversational search methods, the quality of the learned model is heavily dependent on the availability of a large amount of annotated training data. However, obtaining such annotated data can be expensive and time-consuming, as it requires human annotators to engage in multi-turn conversations with the system, label the retrieved passages, rewrite the queries, and provide feedback. Furthermore, the dynamic and contextual nature of conversational search makes it difficult to create a comprehensive and representative dataset that covers all possible user interactions and intents.

Data sparsity issues in conversational search significantly hinder the progress of supervised conversational search methods. To overcome the dependency on the availability of conversational search and human-rewritten query data, a few zero-shot query contextualization dense retrievers have been proposed recently~\cite{yu2021few, krasakis2022zero}. These approaches aim to implicitly incorporate the conversational queries and their corresponding context information into query embeddings via a conversational query encoder, without explicitly addressing the inherent linguistic challenges of raw conversational queries: coreference and omission. While these methods provide an end-to-end solution for conversational search, their limitations are also apparent.





\begin{table}[]

\caption{An conversational search dialogue in TREC CAsT. The coreference ambiguity of $q_4$ and its ZeQR resolution are shown in \textcolor{ForestGreen}{green}. The omission ambiguity of $q_4$ and its ZeQR resolution are shown in \textcolor{red}{red}. The information sources for query reformulation are marked using \underline{underline}.}
\vspace{-2mm}
\resizebox{0.46\textwidth}{!}{%
\begin{tabular}{p{9cm}}
\hline
\textbf{Dialogue context ($t=4$)}                                                                         \\ \hline
$q_1$: I just had a breast biopsy for cancer. What are the most common types?                                                    \\
\multicolumn{1}{l}{\hspace{0.5cm}$a_1$: ...Non-invasive breast cancer is when the cancer is ...}                                                                    \\
$q_2$: Once it breaks out, how likely is it to spread?                                                            \\
\multicolumn{1}{l}{ \hspace{0.5cm}$a_2$: ...How is \underline{Lobular Carcinoma in Situ} diagnosed? You often...}                                                                    \\
$q_3$: How deadly is \underline{Lobular Carcinoma in Situ}?                                                    \\
\multicolumn{1}{l}{\hspace{0.5cm}$a_3$: ...In this case it will be described as \underline{Lobular Neoplasia}...}                                                                    \\
$q_4$: Wow, \textcolor{ForestGreen}{that} is better than I thought.  What are common \textcolor{red}{treatments}?\vspace{0.1cm}                                                               \\ \hhline{=}
\textbf{ZeQR resolution}                                                                            \\ \hline
$q_4^{**}$: Wow, \textcolor{ForestGreen}{Lobular Neoplasia} is better than I thought.  What are common \textcolor{red}{treatments of Lobular Carcinoma in Situ}? \vspace{0.1cm} \\ \hline

\end{tabular}%
}
\label{tab:example}
\vspace{-3mm}
\end{table}



First, the existing methods are not universally applicable to arbitrary retrievers. For each newly introduced dense retrieval model, ad-hoc modifications are necessary to accommodate conversational input. Also, a successful adaptation demands significant computational resources and storage space for experiment and indexing. Second, the explainability of query contextualization methods is low, as they attempt to address conversational search through the internal operations of dense query encoders. This makes the investigation of their effectiveness very difficult. Lastly, the ability to resolve conversational search queries via contextualization heavily relies on the attention mechanism of Transformer encoders. As a result, they face difficulties in handling ambiguities caused by omission.

To address these limitations, in this paper, we propose a novel Zero-shot Query Reformulation (ZeQR) framework that eliminates the need for conversational search data such as rewritten queries or relevance judgments, rendering it a zero-shot approach.




Unlike existing query contextualization methods that depend on specific dense retrieval models, the ZeQR framework is method-agnostic, allowing for the seamless adaptation of arbitrary dense retrieval methods to conversational search without the need for additional indexing or retraining. This versatility not only broadens the applicability of our framework but also directly bridges the development of new retrieval methods to the conversational search domain.


Furthermore, the effectiveness of current query contextualization methods in conversational search is primarily attributed to the bi-directional attention mechanism of the Transformer encoder architecture. While this mechanism allows them to handle coreference ambiguities relatively well, it also presents challenges when faced with ambiguities caused by omission, because resolving omission involves appending additional information to words in addition to altering information. In contrast, ZeQR proactively addresses the omission and coreference resolutions by explicitly formulating specific tasks within the framework, simultaneously enhancing its explainability.

Despite the benefits of zero-shot query reformulation over existing zero-shot query contextualization techniques, constructing an effective query reformulator without supervision from conversational search data remains a significant challenge. Currently, to the best of our knowledge, all query reformulators rely on the direct supervision of conversational data, and our research is the first to tackle this challenge.

The proposed zero-shot Query Reformulation (ZeQR) framework provides a unified solution to tackle both coreference and omission resolution. Table~\ref{tab:example} shows an example of the resolution results of ZeQR. To evaluate the effectiveness of ZeQR, we conducted comprehensive experiments on four widely used conversational datasets \cite{2019trec,2020trec, 2021trec, 2022trec}. The experimental results demonstrate that our method consistently outperforms existing state-of-the-art zero-shot baselines, emphasizing its effectiveness in enhancing conversational search performance. Further analysis revealed that the major source of this advantage was attributed to the omission ambiguity resolution task.





\section{Related Work}

Recently, several zero-shot query contextualization methods have been proposed to address zero-shot conversational search, namely ConvDR (Zero-shot)~\cite{yu2021few} and ZeCo\textsuperscript{2}~\cite{krasakis2022zero}. In ConvDR, the teacher model is an ANCE~\cite{xiong2020ance} with frozen parameters, which leverages labeled query-document pairs to provide supervision to the student model. The student model's parameters are learnable and initialized using the ANCE checkpoint, where the student model does not require manually reformulated queries as input. For its zero-shot run, the authors employ the student model without supervision from conversational search datasets. ZeCo\textsuperscript{2}, the current state-of-the-art method for zero-shot conversational retrieval, is built on ColBERT~\cite{khattab2020colbert}. ZeCo\textsuperscript{2} obtains query contextualized embeddings by truncating the output of the ColBERT encoder. Specifically, unlike ColBERT, which calculates relevance scores through exhaustive MaxSim operations between all embeddings from a query and a document, ZeCo\textsuperscript{2} only uses contextualized embeddings from the encoder that correspond to the query's positions.


Both ConvDR (Zero-shot) and ZeCo\textsuperscript{2} address zero-shot conversational search by implicitly relying on specific dense retrieval models. This dependence hinders their ability to take advantage of newly-developed dense retrieval methods, potentially limiting the versatility and adaptability of these approaches. Furthermore, while the bidirectional architecture of Transformer encoders allows them to handle coreference ambiguities relatively well by capturing context and bringing the embeddings of pronouns closer to their anaphora through the assignment of larger attention weights to the actual referent rather than the pronoun~\cite{krasakis2022zero}, this characteristic also presents challenges in resolving omission ambiguities. This is because omission ambiguities involve appending additional information to words, rather than merely altering the existing information. Consequently, the attention mechanism of Transformer encoders, which primarily focuses on adjusting attention weights, may struggle to effectively resolve these omissions. On the contrary, ZeQR provides a unified solution to tackle both coreference and omission resolution explicitly. By reformulating conversational queries into context-independent queries containing relevant information, query reformulation converts conversational search into ad-hoc search, obviating the need for extensive modifications of existing dense retrievers to support conversational search.

\begin{figure*}[h]
  \centering
  \includegraphics[scale=0.3]{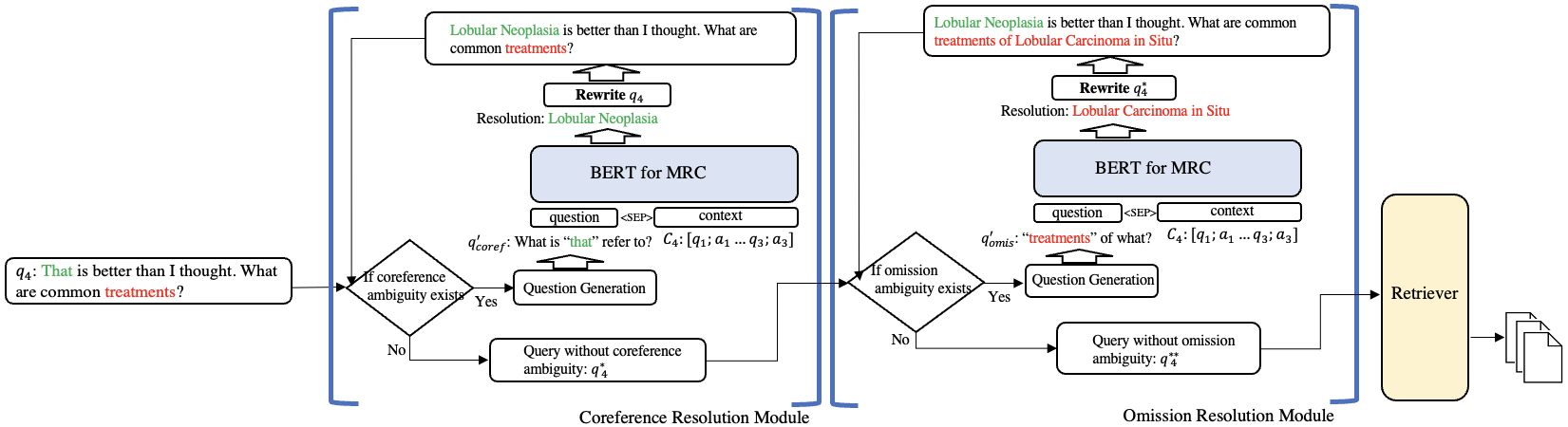}
  \caption{Overview of our proposed ZeQR. Showing the query reformulation process of the same query example of Table~\ref{tab:example}}
  \label{fig:method}
\end{figure*}

\section{Methodology}

In this section, we show how we solve the zero-shot query reformulation problem by converting the original task to machine reading comprehension problems, and how ZeQR becomes a unified solution to tackle both coreference and omission resolution tasks in conversational search.

\subsection{Task Definition and Notation}

A conversational retrieval system aims to find relevant passages for each given query in a conversational search session $\mathbf{S}=\{\langle q_1, a_1 \rangle,\\ \langle q_2, a_2 \rangle, \cdots, \langle q_N, a_N \rangle\}$, where a session has $N$ consecutive turns $\langle q_t, a_t \rangle$. $q_t$ and $a_t$ are the raw query from a user and the corresponding canonical passage at the t-th turn respectively. For each turn, the conversational retrieval system retrieves relevant passages with regard to query $q_t$ and its context $\mathbf{C_t} \subseteq \{q_{<t}, a_{<t}\}$, where $q_{<t}$ and $a_{<t}$ stand for all the user-initiated questions, and their corresponding canonical answers before the $t_{th}$ step respectively. However, directly using the raw queries $q_t$ or the concatenation of $q_t$ and $\mathbf{C_t}$ on retrieval systems usually leads to sub-optimal results. The task of query reformulation in conversational search is to de-contextualize the raw query $q_t$ based on the context $\mathbf{C_t}$ and generate a context-independent query $\acute{q}_t$.

\subsection{A Two-step Framework}

The difficulty of query reformulation in conversational search fundamentally stems from two linguistic phenomena in conversations: coreference and omission~\cite{yu2020few, lin2021multi}. 

\begin{itemize}
    \item Coreference: replacing complex terminologies mentioned in context with pronouns.
    \item Omission: the description of a verb or noun is omitted if the description has been mentioned in context. The description can be an adjective or a prepositional phrase.
\end{itemize}

Therefore, a query reformulation task can be divided into two sub-problems: coreference resolution and omission resolution for the focal query, given the same context information. 

Ideally, both sub-problems should be effectively handled by a unified framework. Hence, a potential candidate model should maintain flexibility in the problem formulation to accommodate different input formats of different sub-problems.
Moreover, we aim to transit data dependency from data in conversational search domain to a more general domain with abundant data resources available, in order to reduce the long-term costs associated with improving conversational search performance.

Consequently, Machine reading comprehension (MRC) is a promising area for this purpose, as it offers a highly flexible question format and a wealth of large-scale datasets such as SQuAD, SQuAD2, and QuAC~\cite{rajpurkar2016squad, rajpurkar2018know, choi2018quac}. More importantly, MRC models can often provide accurate coreference and omission resolutions when posed with the right questions.

Thus, we propose the ZeQR framework, which leverages existing MRC language models to answer a set of template-based, well-designed questions directly targeting the two aforementioned ambiguities in conversational search. The framework consists of two steps, with each step designed to tackle one ambiguity: coreference resolution and omission resolution. Figure~\ref{fig:method} shows the overview of the proposed ZeQR framework.

In summary, the ZeQR framework uses MRC language models to identify and resolve ambiguities in queries. By reformulating the query based on the outputs of the MRC models, it effectively converts the query reformulation task into an MRC problem, thus automatically determining the most relevant referent or descriptive information in the context.



\subsection{Coreference Resolution}

To achieve coreference resolution through machine reading comprehension, a universal template is employed to generate a coreference resolution question $q'_{coref}$ for each sample, which serves as the main input for a machine reading comprehension model.

To identify coreferences in a query $q$, our framework first checks if a pronoun is present. If so, it generates a corresponding question $q'_{coref}$ using the template: \textit{What is [ ] refer to, in ``\{ \}"}. The pronoun is inserted into the first blank \textit{[ ]}, and the original query $q$ is inserted into the second blank \textit{\{ \}}. The machine reading comprehension pipeline then takes the formulated question $q'_{coref}$ and combines it with the context $\mathbf{C}$ to form the input. The answer from the pipeline is expected to be the referent of the pronoun, which is then used to replace the pronouns in the original query $q$ to generate the reformulated query $q^{*}$.

\subsection{Omission Resolution}

To improve the efficiency of the omission resolution process, the module only resolves omission resolution for important words with IDF(Inverse Document Frequency) scores higher than a hyperparameter $\eta$. If an important noun or verb does not have an adjective phrase preceding it or a prepositional phrase following it, an automatically generated question $q'_{omis}$ will be formulated from a universal template: \textit{[ ] ( ) what, in ``\{ \}"}. In the template, the pronoun is placed in the first blank \textit{[ ]}, a preposition (either \textit{of} for noun phrases or \textit{to} for verbs) is placed in the second blank \textit{( )}, and the query $q^{*}$ is placed in the last blank \textit{\{ \}}.

The answer from the MRC pipeline is expected to be a description of the focal verb or noun phrase. In the reformulated query $q^{**}$, the descriptions will be added to the corresponding noun phrases and verbs. By these two steps, the proxy of $\acute{q}$, which is the reformulated query $q^{**}$, will be passed on to the retrieval model.

\begin{table*}
  \caption{Overall effectiveness of Zero-shot query reformulation on TREC CAsT. Superscripts ${ }^{\ddagger}$ and ${ }^{\P}$ indicate statistically significant improvements over ConvDR (Zero-shot) ${ }^{\ddagger}$ and ZeCo\textsuperscript{2} ${ }^{\P}$ (paired t-test, $\text{p value} < 0.05$) respectively. The best-performing results from zero-shot retrieval methods in each group are marked in Bold.}
  \label{tab:performance_comparison}
\begin{tabular}{lcllllllll}
\hline
\multicolumn{2}{c|}{\multirow{2}{*}{\textbf{}}} & \multicolumn{4}{c|}{\textbf{CAsT-19}}                                                                     & \multicolumn{4}{c}{\textbf{CAsT-20}}                                                      \\ \cline{3-10} 
\multicolumn{2}{c|}{}                                 & \textbf{NDCG@5}      & \textbf{P@5}         & \textbf{R@100}       & \multicolumn{1}{l|}{\textbf{MAP}}    & \textbf{NDCG@5}      & \textbf{P@5}         & \textbf{R@100}       & \textbf{MAP}         \\ \hline
\textbf{Methods}   & \textbf{zero-shot}             & \multicolumn{1}{l}{} & \multicolumn{1}{l}{} & \multicolumn{1}{l}{} & \multicolumn{1}{l}{}                 & \multicolumn{1}{l}{} & \multicolumn{1}{l}{} & \multicolumn{1}{l}{} & \multicolumn{1}{l}{} \\ \hline
ConvDR-ZS               & \multicolumn{1}{c|}{\cmark}  &  0.2400               & 0.3052               & 0.1825               & \multicolumn{1}{l|}{0.0962}          & 0.1389               & 0.1635               & 0.1508               & 0.0613               \\
ZeCo\textsuperscript{2}                & \multicolumn{1}{c|}{\cmark}  & 0.2306               & 0.3052               & 0.2155               & \multicolumn{1}{l|}{0.1111}          & 0.1615               & 0.1971               & 0.2003               & 0.0864               \\

ZeQR (ours)          & \multicolumn{1}{c|}{\cmark}          & \textbf{0.3821}${ }^{\ddagger}$${ }^{\P}$      & \textbf{0.4751}${ }^{\ddagger}$${ }^{\P}$      & \textbf{0.3575}${ }^{\ddagger}$${ }^{\P}$               & \multicolumn{1}{l|}{\textbf{0.2095}${ }^{\ddagger}$${ }^{\P}$} & \textbf{0.2281}${ }^{\ddagger}$${ }^{\P}$      & \textbf{0.2981}${ }^{\ddagger}$${ }^{\P}$      & \textbf{0.3116}${ }^{\ddagger}$${ }^{\P}$      & \textbf{0.1462}${ }^{\ddagger}$${ }^{\P}$      \\ \hline

\textbf{Reference}   & \multicolumn{1}{l}{}           & \multicolumn{1}{l}{} & \multicolumn{1}{l}{} & \multicolumn{1}{l}{} & \multicolumn{1}{l}{}                 & \multicolumn{1}{l}{} & \multicolumn{1}{l}{} & \multicolumn{1}{l}{} & \multicolumn{1}{l}{} \\ \hline

T5 Rewriter          & \multicolumn{1}{c|}{\xmark}          & 0.2662               & 0.3908               & 0.3995               & \multicolumn{1}{l|}{0.2268}          & 0.2855               & 0.3721               & 0.3791               & 0.1899               \\
Manual               & \multicolumn{1}{c|}{\xmark}           & 0.4056               & 0.5087               & 0.3626               & \multicolumn{1}{l|}{0.2014}          & 0.4121               & 0.4894               & 0.4082               & 0.2246               \\ \hhline{==========}

\multicolumn{2}{c|}{\multirow{2}{*}{\textbf{}}} & \multicolumn{4}{c|}{\textbf{CAsT-21}}                                                                     & \multicolumn{4}{c}{\textbf{CAsT-22}}                                                      \\ \cline{3-10} 
\multicolumn{2}{c|}{}                                 & \textbf{NDCG@5}      & \textbf{P@5}         & \textbf{R@100}       & \multicolumn{1}{l|}{\textbf{MAP}}    & \textbf{NDCG@5}      & \textbf{P@5}         & \textbf{R@100}       & \textbf{MAP}         \\ \hline
\textbf{Methods}   & \textbf{zero-shot}             & \multicolumn{1}{l}{} & \multicolumn{1}{l}{} & \multicolumn{1}{l}{} & \multicolumn{1}{l}{}                 & \multicolumn{1}{l}{} & \multicolumn{1}{l}{} & \multicolumn{1}{l}{} & \multicolumn{1}{l}{} \\ \hline
ConvDR-ZS               & \multicolumn{1}{c|}{\cmark}  & 0.1920               & 0.2633               & 0.2679               & \multicolumn{1}{l|}{0.1220}          & \multicolumn{1}{l}{0.0744} & \multicolumn{1}{l}{0.1117} & \multicolumn{1}{l}{0.0805} & \multicolumn{1}{l}{0.0369} \\

ZeCo\textsuperscript{2}                & \multicolumn{1}{c|}{\cmark}  & 0.2216               & 0.2987               & 0.2670               & \multicolumn{1}{l|}{0.1203}          & N/A                    & N/A           & N/A                    & N/A                    \\
ZeQR (ours)          & \multicolumn{1}{c|}{\cmark}          & \textbf{0.2712}${ }^{\ddagger}$${ }^{\P}$      & \textbf{0.3316}${ }^{\ddagger}$      & \textbf{0.3453}${ }^{\ddagger}$${ }^{\P}$               & \multicolumn{1}{l|}{\textbf{0.1720}${ }^{\ddagger}$${ }^{\P}$} & \textbf{0.1998}${ }^{\ddagger}$      & \textbf{0.2687}${ }^{\ddagger}$      & \textbf{0.1527}${ }^{\ddagger}$               & \textbf{0.0956}${ }^{\ddagger}$      \\ \hline
\textbf{Reference}   & \multicolumn{1}{l}{}           & \multicolumn{1}{l}{} & \multicolumn{1}{l}{} & \multicolumn{1}{l}{} & \multicolumn{1}{l}{}                 & \multicolumn{1}{l}{} & \multicolumn{1}{l}{} & \multicolumn{1}{l}{} & \multicolumn{1}{l}{} \\ \hline
T5 Rewriter          & \multicolumn{1}{c|}{\xmark}           & 0.3058               & 0.3899               & 0.3838               & \multicolumn{1}{l|}{0.2042}          & 0.2226               & 0.3067               & 0.1779               & 0.1064               \\
Manual               & \multicolumn{1}{c|}{\xmark}          & 0.4047               & 0.4747               & 0.4027               & \multicolumn{1}{l|}{0.2178}          & 0.3856               & 0.5067               & 0.3099               & 0.1890               \\ \hline
\end{tabular}
\end{table*}

\section{Experimental Setup}

\subsection{Datasets and Metrics}

To evaluate our proposed method, we conduct experiments on the TREC CAsT-19, CAsT-20, CAsT-21, and CAsT-22 datasets~\cite{2019trec, 2020trec, 2021trec, 2022trec}. The CAsT-19 dataset comprises 50 conversational search sessions, whereas the remaining datasets each contain approximately 26 sessions per year. Each session consists of around 10 consecutive turns. Notably, in CAsT-20, CAsT-21, and CAsT-22, a canonical passage is provided for each search turn, enabling users to offer feedback on these passages.

Canonical passages, which are generally much longer than queries, introduces greater challenges in accurately identifying reference and omission resolutions within these three datasets compared to CAsT-19. To assess the effectiveness of both the baselines and our proposed method, we employ four metrics: NDCG@5, Precision@5, MAP, and Recall@100. In accordance with the official TREC CAsT overview, NDCG@5 serves as our primary metric.

\subsection{Baselines}

In order to evaluate the efficacy of our zero-shot query reformulation method, ZeQR, we have compared it against the following baselines:

\begin{itemize}
    \item ConvDR-ZS~\cite{yu2021few}: this refers to the ConvDR (Zero-shot) method. ConvDR is originally developed for few-shot query contextualization through knowledge distillation on query rewrites, but it can also be applied in a zero-shot setting without supervision from conversational tasks or data.
    \item ZeCo\textsuperscript{2}~\cite{krasakis2022zero}: a state-of-the-art zero-shot conversational search approach specifically tailored to address the zero-shot conversational search problem using a modified ColBERT encoder.
\end{itemize}

 For reference purposes, we include the retrieval performance of two non-zero-shot methods:
\begin{itemize}
    \item T5 Rewriter: a widely-used supervised neural query reformulator based on the sequence-to-sequence generative language model, T5~\cite{lin2021multi}. It is trained on a conversational query reformulation dataset, CANARD~\cite{elgohary2019can}.
    \item Manual: the manual rewritten queries provided by TREC.
\end{itemize}

\subsection{Implementation Details}

\subsubsection{Query Reformulation}

 One thing worth to be mentioned is that the order of the two-step framework is important, as pronoun replacement may introduce new omission ambiguities into a query.\footnote{The source code and the experiments can be found at \url{https://github.com/dayuyang1999/ZeQR}} To illustrate, consider the sample in Table 1. If $a_3$ had mentioned the word "treatment" at the end, the user may have asked "That is better than I thought. What are common \underline{ones}" instead of "That is better than I thought. What are common \underline{treatments}". In this case, if the coreference ambiguity is not resolved first, the omission ambiguity with respect to "treatment" will remain unsolved.

To convert two subproblems into MRC problems, the input of MRC is formulated as the concatenation of a question and its corresponding context, with a special token <SEP> placed in the middle to help the model identify their positions. The MRC model then extracts the answer solely from the context segment, which fits the setting of conversational search since the referents and descriptions required will only arise from the context.

To keep the simplicity of our method, we fine-tuned a BERT on a standard MRC dataset: SQuAD~\cite{rajpurkar2016squad}. However, using BERT~\cite{devlin2018bert} as the basic model of the two resolution modules in ZeQR has a drawback due to the restriction on input token length of BERT. Instead of employing a rolling window to truncate input, we simply consider the most recent canonical passage. This is because during a conversational search session, the user usually asks questions and the system retrieves long passages to meet their information needs. Intuitively, the user is more likely to anaphorize a word shown in the most recent passage if they are about to ask a question about it, or anaphorize a word from their previous questions to further explore the topic. Additionally, an experiment on CAsT-21~\cite{dayu2022trec} shows that the best-performing reformulated queries only take the most recent canonical passage into account. Consequently, we limit ourselves to the most recent passage. If the overall input length surpasses the limit of BERT, we truncate the canonical passage (which happened only once across all four CAsT datasets).

For training the BERT for MRC, we use the commonly-used SQuAD~\cite{rajpurkar2016squad} dataset for which we follow the original split, using 87599 samples for training and 10570 samples for validation. We use bert-base-cased\footnote{https://huggingface.co/bert-base-cased} checkpoint as our base model. Although we do not use the rolling window technique during inference, to handle the examples in the SQuAD dataset having very long contexts, we use the rolling window technique during training, setting the maximum possible length for context to be 384 (not 512 to save some spaces for the question) and 50 tokens as the size of the rolling window. We use Weight-decay Adam as our optimizer, with a learning rate of 2e-5 and a linear weight decay function with a decay rate of 0.01. 

During inference, the threshold $\eta$ for identifying important words was set to 2.65. In addition, we always use BM25 parameter settings as k1=0.9 and b=0.4.

\subsubsection{Retriever}

We use TCT-ColBERT~\cite{lin2020tctcolbert} as the follow-up retriever after query reformulation. The reason we choose TCT-ColBERT over ColBERT~\cite{khattab2020colbert} is due to the latter's prohibitive memory usage. While TCT-ColBERT employs knowledge distillation techniques to achieve comparable performance to ColBERT with substantially reduced computational cost. We implement the follow-up retrievers with an open-sourced Python toolkit, Pyserini\footnote{https://github.com/castorini/pyserini}.

\subsubsection{Baselines}

For computing the metrics for ConvDR (Zero-shot) on CAsT-19 and CAsT-20, we use the run files generously provided by the authors of ZeCo\textsuperscript{2}. For CAsT-21 and CAsT-22, we re-implement ConvDR's Zero-shot run based on the official release of ConvDR. However, instead of using the original ANCE checkpoint from Microsoft\footnote{https://github.com/microsoft/ANCE}, we use an ANCE checkpoint from huggingface\footnote{https://huggingface.co/castorini/ance-msmarco-passage}, as the original checkpoint is no longer public available.

For ZeCo\textsuperscript{2}, we obtain the metrics directly from the run files provided by the authors of ZeCo\textsuperscript{2}. Unfortunately, we do not get the run file for CAsT-22, and the memory consumption of ColBERT is gigantic for CAsT-22 if we follow the same indexing setting as the authors did (ColBERT version=1, FAISS sampling ratio=30\%) because the collection size of CAsT-22 is over 4x larger than the other three CAsT datasets. As a result, 
we do not report the results of ZeCo\textsuperscript{2} for CAsT-22, which ends with some N/As in Table~\ref{tab:performance_comparison}.

\section{Results and analysis}

In this section, we conduct a series of experiments to evaluate the effectiveness of the ZeQR framework and investigate the sources of its performance advantage.

\subsection{Effectiveness of ZeQR}

We first performed a comparative analysis to assess the effectiveness and improvements of ZeQR over two established zero-shot query contextualization baselines. This evaluation was conducted on four conversational search datasets: TREC CAsT-19, CAsT-20, CAsT-21, and CAsT-22, as shown in Table~\ref{tab:performance_comparison}. 

Our findings demonstrate that ZeQR outperforms all existing zero-shot query contextualization methods across a variety of metrics, including NDCG, Precision (P), Recall (R), and Mean Average Precision (MAP). And the performance gap between ZeQR and the baselines is statistically significant in most cases.

In the CAsT-19 dataset, ZeQR demonstrates retrieval effectiveness that is nearly on par with manually rewritten queries and even exceeds the performance of the T5 rewriter. This result is particularly surprising, considering the T5 rewriter is built by fine-tuning a powerful generative language model using human-annotated query reformulation data. Furthermore, ZeQR maintains performance levels comparable to the T5 query rewriter in the CAsT-20, CAsT-21, and CAsT-22 datasets.

\begin{table*}[]
\begin{tabular}{p{3cm}p{4cm}|p{0.7cm}p{0.7cm}p{0.7cm}}
\hline

\multicolumn{1}{c}{\multirow{2}{*}{Raw   Utterance}}   & \multicolumn{1}{c|}{\multirow{2}{*}{ZeQR resolution}}    & \multicolumn{3}{c}{NDCG@5}                                                               \\ \cline{3-5}

\multicolumn{1}{c}{}                       & \multicolumn{1}{c|}{}                                    & \multicolumn{1}{c}{ConvDR-ZS} & \multicolumn{1}{c}{ZeCo\textsuperscript{2}} & \multicolumn{1}{c}{ZeQR} \\

\hhline{=====}

What is the \textcolor{red}{difference} with Bologna?                &  What is the \textcolor{red}{difference of mortadella} with Bologna?      & 0                          & 0                         & 0.800                          \\ \hline
What are the \textcolor{red}{EU rules}?                               &  What are the \textcolor{red}{EU rules of GMO Food labeling}?              & 0                          & 0                         & 0.744                          \\ \hline
What \textcolor{red}{licenses and permits} are needed?                &  What \textcolor{red}{licenses and permits of food truck} are needed? & 0                          & 0                         & 0.889                           \\ \hline
What is its main \textcolor{red}{economic activity}?                  & What is its main \textcolor{red}{economic activity of Salt Lake City}?    & 0.111                    & 0                         & 0.875                          \\ \hline
\end{tabular}
\caption{Examples of TREC CAsT samples that have the largest performance gap between ZeQR and baseline methods. The ambiguity in the raw utterance and corresponding resolutions are marked as \textcolor{red}{red}.}
\label{tab:compare2}
\end{table*}

\begin{figure}
    \centering
    \includegraphics[scale=0.23]{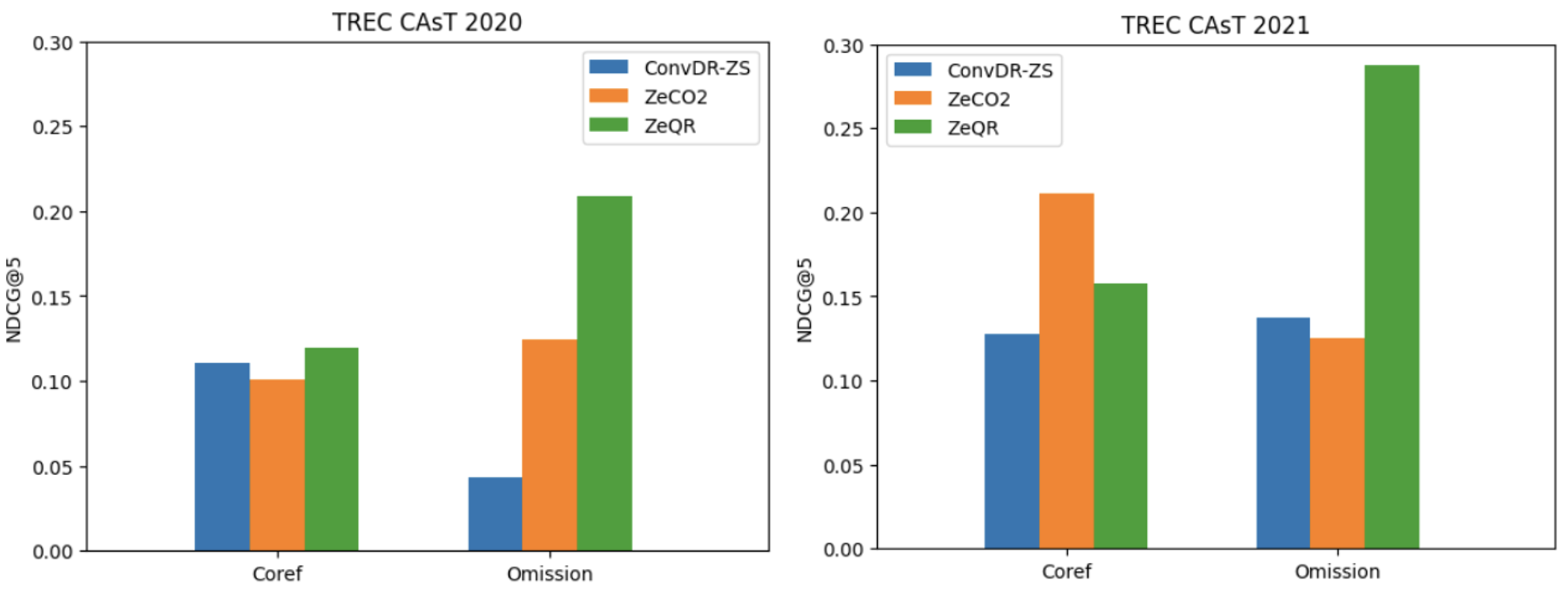}
    \caption{Retrieval performance comparison of \textcolor{RoyalBlue}{ConvDR-ZS}, \textcolor{orange}{ZeCo\textsuperscript{2}}, and \textcolor{ForestGreen}{ZeQR}, on samples with either only coreference(coref) or omission ambiguity.}
    \label{fig:omis_adv}
\end{figure}

\subsection{Source of Performance Advantage: Omission Resolution Task}

\begin{table}[]
\begin{tabular}{l|c|c}
\hline
            & \multicolumn{1}{l|}{\textbf{CAsT-20}} & \multicolumn{1}{l}{\textbf{CAsT-21}} \\ \hline
Coreference & 61                             & 66                            \\ \hline
Omission    & 96                             & 112                           \\ \hline
\end{tabular}
\caption{Number of raw queries having omission or coreference ambiguities.}
\label{table:stat_omis_coref}
\end{table}

Our comprehensive experiments conducted on multiple datasets demonstrate the superior performance of ZeQR in comparison to existing zero-shot query contextualization methods. To better understand this performance advantage, we aim to identify the scenarios that contribute most to the performance improvement. Therefore, we aim to compare samples with the largest performance gap between ZeQR and the baselines. Specifically, we identified and collected the samples with a performance gap exceeding a significant threshold, 0.7, on our primary evaluation metric, NDCG@5, between ZeQR and the baseline methods. The samples we found are shown in~Table~\ref{tab:compare2}.

 We observe that all these cases can be categorized into a single scenario: raw utterances containing omission ambiguity. This observation suggests that the omission resolution capability of existing zero-shot query contextualization methods is not effective. One possible reason is that the attention mechanism can only modify an embedding's value by assigning different attention weights, whereas omission resolution necessitates the addition of information in the context. Moreover, the improved explainability of ZeQR over baseline methods enables us to directly analyze these examples by examining their linguistic formulation. We find that ZeQR successfully appends specific descriptions to their corresponding words, which explicitly resolves omission ambiguity and leads to enhanced retrieval performance.

To further validate our hypothesis that ZeQR's performance advantage mainly stems from its superior ability to resolve omission ambiguities, we collected samples containing only coreference or omission ambiguities from the TREC CAsT-20 and CAsT-21 datasets\footnote{TREC CAsT-19 does not have canonical passages provided.}. We evaluated the retrieval performance of ZeQR and two baseline methods using NDCG@5. As depicted in Figure~\ref{fig:omis_adv}, while the performance of all three methods on queries with coreference ambiguities is similar, the performance on queries containing only omission ambiguities varies, with ZeQR significantly outperforming ZeCo\textsuperscript{2} and ConvDR-ZS.

Considering ZeQR's superior performance on omission tasks, we further explored the importance of omission tasks on overall performance by counting the number of occurrences of both ambiguities in the CAsT-20 and CAsT-21 datasets. As Table~\ref{table:stat_omis_coref} shows, the number of occurrences of omission ambiguity in original queries surpasses that of coreference ambiguity, indicating that omission ambiguity is as prevalent as coreference ambiguity in raw queries.  Therefore, improving the omission resolution task can greatly contribute to enhancing conversational search.

\begin{figure}
    \centering
    
    \includegraphics[scale=0.15]{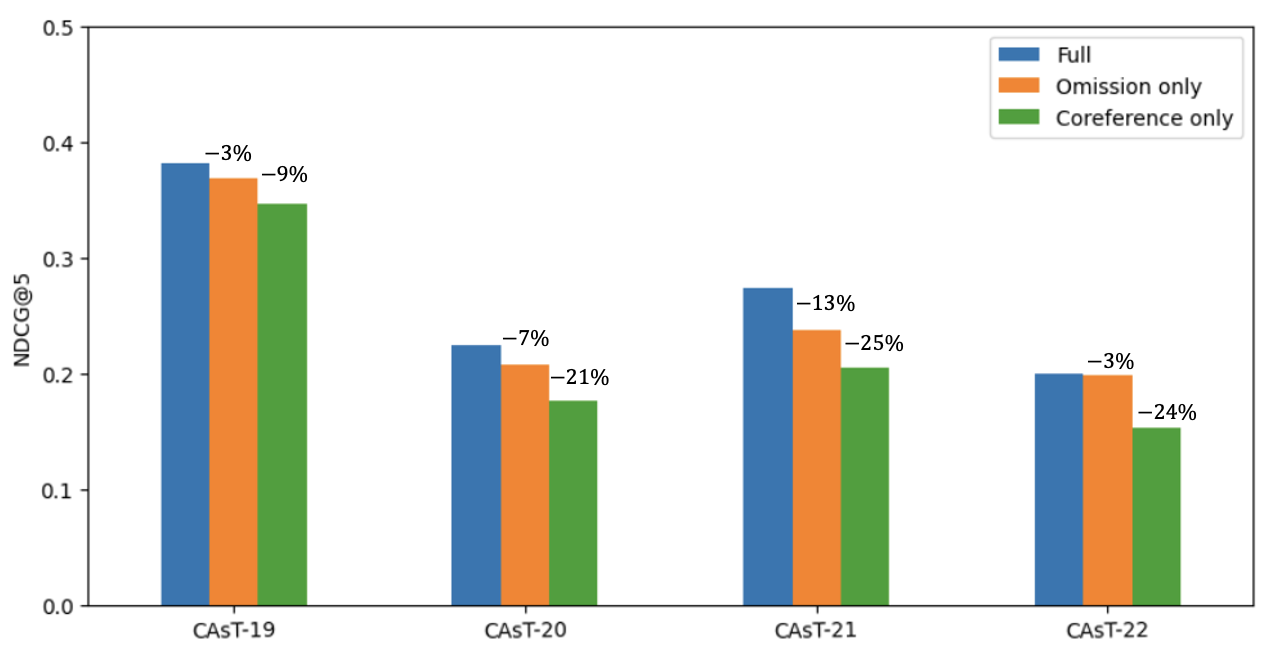}
\caption{Relative performance drops comparing full ZeQR model with coreference only or omission only model.}
    \label{fig:ablation}
    \vspace{-0.4cm}
\end{figure}

\subsection{Ablation Study}

In this section, we present an ablation study to assess the individual contributions of the coreference and omission resolution components of ZeQR to retrieval performance. We evaluated the impact of removing either the coreference (omission only) or omission (coreference only) resolution module. The results are depicted in Figure~\ref{fig:ablation}, demonstrating that eliminating the omission module led to a larger impact, with an average decrease of 20\% on NDCG@5, whereas removing the coreference module resulted in an average decrease of 6\% on NDCG@5. Regardless of the dataset, the full two-step framework outperforms the models with either component removed.

Additionally, Figure~\ref{fig:ablation} reveals that the influence of removing the omission ambiguity resolution task may be even more pronounced when canonical passages are considered. Comparing the CAsT-19 (where canonical passages were not included) and the CAsT-22 (which incorporates canonical passages), the performance drop for the "omission only" model remains the same at 3\%. However, the performance decline for the "coreference only" model has risen from 9\% on CAsT-19 to 25\% on CAsT-22. This suggests omission resolution is particularly important especially when canonical passages are considered.

\begin{table}[]
\begin{tabular}{l|llll}
\hline
\multicolumn{1}{c|}{\multirow{2}{*}{\textbf{\begin{tabular}[c]{@{}c@{}}ZeQR's   follow-up\\       Retriever\end{tabular}}}} & \multicolumn{1}{l|}{\textbf{CAsT19}} & \multicolumn{1}{l|}{\textbf{CAsT20}} & \multicolumn{1}{l|}{\textbf{CAsT21}} & \textbf{CAsT22} \\ \cline{2-5} 
\multicolumn{1}{c|}{}                                                                                                     & \multicolumn{4}{c}{\textbf{NDCG@5}}                                                                                                      \\ \hline
TCT-ColBERT                                                                                                               & 0.3821                                & 0.2281                                & 0.2712                                & 0.1998 
           \\
ANCE                                                                                                                      & 0.3008                                & 0.1734                                & 0.2450                                & 0.1440           \\
BM25                                                                                                                       & 0.2605                                  & 0.1333                                  & 0.1954                                 &    0.1169
          \\ \hline
\end{tabular}
\caption{Impact of using different retrievers on the overall performance of ZeQR}
\label{tab:general}
\end{table}

\subsection{Impact of Using Different Retriever}

The generality of the ZeQR framework enables the integration of various retrievers as a subsequent retrieval step, thereby allowing us to assess the impact of employing different retrievers on zero-shot conversational search. Table~\ref{tab:general} demonstrates that TCT-ColBERT is the most effective retriever for conversational search tasks. This result is expected, as resolving omission and coreference ambiguities requires a fine-grained understanding of queries at the word level. TCT-ColBERT achieves this by mimicking ColBERT's word-to-word matching function, while maintaining lower resource consumption. In comparison, another popular dense retriever, ANCE, still delivers competitive performance when measured against the sparse retriever BM25. This observation suggests that comprehending the semantic compatibility between queries and documents may be a crucial capability in conversational search compared with bag-of-words matching.

\section{conclusions and future work}

In this paper, we introduced ZeQR, a novel zero-shot query reformulation framework designed to transform query reformulation tasks into machine reading comprehension problems in conversational search. Our framework surpasses existing zero-shot conversational retrieval methods in terms of efficacy and ease of implementation. Additionally, we highlighted the importance of omission ambiguity resolution, an aspect often overlooked in prior research, and demonstrated our method's substantial performance advantage in this particular task. We anticipate that our findings will further enrich the ongoing discussion surrounding the potential benefits of shifting the dependency of conversational search tasks from conversational search datasets to datasets in a more general domain.

As for future research directions, we aim to explore the impact of employing different MRC datasets on ZeQR's zero-shot retrieval performance. This examination is crucial, as certain question formats in MRC datasets may not be inherently advantageous to the two resolution tasks of our primary interest. Moreover, we plan to investigate strategies for mitigating the computational complexity introduced by ZeQR's two-step resolution process without compromising its accuracy. Ultimately, we believe that our work holds significant promise for advancing the state-of-the-art in zero-shot conversational search.

\section{Acknowledgments}

This research is supported by the graduate fellowship from the Institute for Financial Services Analytics at University of Delaware. We would also like to express our sincere gratitude to the reviewers for their insightful comments and suggestions.

\balance


\bibliographystyle{ACM-Reference-Format}
\bibliography{sigir2023.bib}

\appendix
\section{Appendix}

Our framework is designed with inherent adaptability in mind, enabling effortless integration with an assortment of Machine Reading Comprehension (MRC) and Question Answering (QA) models, including large language models such as ChatGPT. It's crucial to note, however, that the primary objective of our framework is to empower researchers to autonomously harness its full performance potential, negating the need for dependence on external resources. Instead, we have opted for MRC as our choice of implemented language model. Researchers can readily train their MRC models using standard MRC datasets such as the SQuAD dataset on a consumer-level GPU within a span of just one hour.

Nonetheless, I share the results from the experiments that implemented ChatGPT in the Appendix. These results display a level of performance that is impressively proximate to human capabilities.


\begin{table*}[h!]
\caption{The performance comparison for ZeQR framework using different language models (LMs). The best-performing results are marked in Bold.}
\begin{tabular}{lllllllllll}
\hline
                      & \multicolumn{1}{l|}{}       & \multicolumn{3}{c|}{\textbf{CAsT-20}}                                & \multicolumn{3}{c|}{\textbf{CAsT-21}}                                & \multicolumn{3}{c}{\textbf{CAsT-22}}            \\ \cline{3-11} 
                      & \multicolumn{1}{l|}{}       & \textbf{NDCG@5} & \textbf{P@5} & \multicolumn{1}{l|}{\textbf{R@100}} & \textbf{NDCG@5} & \textbf{P@5} & \multicolumn{1}{l|}{\textbf{R@100}} & \textbf{NDCG@5} & \textbf{P@5} & \textbf{R@100} \\ \hline
\textbf{Methods}      & \textbf{LM}              &                 &              &                                     &                 &              &                                     &                 &              &                \\ \hline
\multirow{2}{*}{ZeQR} & \multicolumn{1}{l|}{MRC}    & 0.2281          & 0.2981       & \multicolumn{1}{l|}{0.3116}         & 0.2712          & 0.3316       & \multicolumn{1}{l|}{0.3453}         & 0.1998          & 0.2687       & 0.1527         \\
                      & \multicolumn{1}{l|}{ChatGPT} & \textbf{0.3486}          & \textbf{0.4452}       & \multicolumn{1}{l|}{\textbf{0.3791}}         & \textbf{0.3867}          & \textbf{0.4810}       & \multicolumn{1}{l|}{\textbf{0.4766}}         & \textbf{0.2878}          & \textbf{0.3865}       & \textbf{0.2554}         \\ \hline
\textbf{Reference}    &                             &                 &              &                                     &                 &              &                                     &                 &              &                \\ \hline
manual                & \multicolumn{1}{l|}{-}      & 0.4121          & 0.4894       & \multicolumn{1}{l|}{0.4082}         & 0.4047          & 0.4747       & \multicolumn{1}{l|}{0.4027}         & 0.3856          & 0.5067       & 0.3099         \\ \hline
\end{tabular}
\label{table:chatgpt}
\end{table*}

As evidenced in Table~\ref{table:chatgpt}, the performance of our framework, ZeQR, when integrated with ChatGPT as the underlying language model, approximates human-level proficiency on the CAsT-20 and CAsT-21 datasets. However, the introduction of a dialogue tree in the CAsT-22 dataset brings additional complexity to the task. This dialogue tree's branching structure allows each conversational session to follow a unique topic flow path, thereby amplifying the challenge for the language model in resolving the resulting ambiguities. Hence, there remains a notable disparity between human performance and that of ZeQR equipped with ChatGPT in handling the CAsT-22 dataset.

\end{document}